# Towards symmetric scheme for superdense coding between multiparties

Andrzej Grudka* and Antoni Wójcik**

Faculty of Physics, Adam Mickiewicz University,

Umultowska 85, 61-614 Poznań, Poland

Abstract

Recently Liu, Long, Tong and Li [Phys. Rev. A 65, 022304 (2002)] have proposed a scheme for superdense coding between multiparties. This scheme seems to be highly asymmetric in the sense that only one sender effectively exploits entanglement. We show that this scheme can be modified in order to allow more senders to benefit of the entanglement enhanced information transmission.

PACS number(s): 03.67.-a, 89.70.+c

Quantum dense coding, first proposed by Bennett and Wiesner [1], is one of quantum information processing protocols which is superior to classical ones. It allows a transmission of two bits of classical information by sending one qubit only. It is possible due to non-local properties of the quantum entanglement. Briefly, two parties (sender and receiver) share two qubits in a given Bell state. Because all four Bell states form an orthogonal base they can be discriminated in a single measurement. Moreover, starting from a given Bell state one can achieve any other Bell state, performing only unitary operations on a single particle. Thus the sender performs one of these unitary operations and sends his particle to the receiver. The receiver when possessing both particles can discriminate among four orthogonal states to obtain two bits of classical information. Quantum dense coding was experimentally presented by Mattle et al. [2] in an optical system, and by Fang et al. [3] by the use of NMR techniques.

Recently Liu, Long, Tong and Li [4] have presented the protocol for superdense coding between $N+1$ users with the use of qudits ($d$-dimensional generalization of qubits). In their scheme $N+1$ parties (one receiver and $N$ senders) share initially an $(N+1)$-particle maximally entangled state. Each particle is $d$-dimensional quantum system. Each party has one particle. Explicitly the initial state is

$$|\Psi_{00...0}\rangle = (|00...0\rangle + |11...1\rangle + ... + |d-1,d-1...d-1\rangle)/\sqrt{d} . \quad (1)$$

The first qubit belongs to the receiver and the remaining N ones to senders. In order to send information each sender performs unitary operation on his particle and then sends it to the receiver. Finally the receiver has one of the $d^{N+1}$ maximally entangled, mutually orthogonal $(N+1)$-particle states of the form

$$\left|\Psi_{i_1,i_2,...i_N}^n\right\rangle = \sum_j e^{2\pi i j n/d} |j\rangle \otimes |j \oplus i_1\rangle \otimes ... \otimes |j \oplus i_N\rangle / \sqrt{d} , \quad (2)$$

where $n, j, i_1, i_2, ...i_N = 0,1,...,d-1$ and $\oplus$ denotes addition modulo $d$. Because each of $n, i_1, i_2, ...i_N$ can take d values, the number of these states is clearly $d^{N+1}$. Because of their orthogonality they can be discriminated by the receiver in a single measurement. Thus he obtains $\log_2 d^{N+1}$ bits of information. The unitary operations which transform between the initial state (1) and the final states (2)

$$U_{i_1,i_2,...i_N}^n = \left|\Psi_{i_1,i_2,...i_N}^n\right\rangle\left\langle\Psi_{00...0}\right| \quad (3)$$

can be obtained as a tensor product

$$U_{i_1,i_2,...i_N}^n = I \otimes U_{i_1}^{n_1} \otimes U_{i_2}^{n_2} \otimes ... \otimes U_{i_N}^{n_N} \quad (4)$$

(where $n = n_1 + n_2 + ... + n_N$) of one-qudit unitary operations

$$U_{i_k}^{n_k} = \sum_{j,j'} e^{2\pi i j n_k / d} \, \delta_{j' j \oplus i_k} |j'\rangle\langle j|. \tag{5}$$

For a given $n$ one can, however, choose $n_1, n_2, ..., n_N$ in many different ways. So in order to avoid confusion, each sender must be obliged to choose his $n_k$ from a given set $S_k$ only. These sets must be constructed in a way allowing only one choice of $n_1, n_2, ..., n_N$ ($n_k \in S_k$) fulfilling the condition $n = n_1 + n_2 + ... + n_N$ for any value of $n$ ($n = 0, 1, ..., d-1$). This restriction is necessary for a reconstruction of each sender's unitary operations by the receiver. Let $|S_k|$ denotes the number of elements of $S_k$. The $k$-th sender is thus allowed to perform one of $|S_k|d$ unitary operations $U_{i_k}^{n_k}$, which is equivalent to a transmission of $\log_2 |S_k| d$ bits of information.

For example, although the authors of [4] did not express it explicitly, they considered a specific construction of sets $S_k$. The set $S_1 = \{0,1,...,d-1\}$ and all other sets $S_{k \neq 1} = \{0\}$. Thus the first sender can transmit $\log_2 d^2$ bits of information but all the remaining senders can transmit only $\log_2 d$ bits of information each. It should be emphasized that the same task can be easily performed without the necessity of $(N+1)$-particle entanglement. One can use only 2 particles in the entangled state and the product state of $N-1$ particles. More explicitly the $(N+1)$-particle state can be of the form

$$|\Phi_{00...0}\rangle = \left(\sum_j |j\rangle \otimes |j\rangle\right) \otimes |0\rangle \otimes ... \otimes |0\rangle / \sqrt{d}. \tag{6}$$

This state can be transformed into one of the mutually orthogonal states

$$|\Phi_{i_1,i_2,...i_N}^n\rangle = \left(\sum_j e^{2\pi i j n / d} |j\rangle \otimes |j \oplus i_1\rangle\right) \otimes |i_2\rangle \otimes ... \otimes |i_N\rangle / \sqrt{d} \tag{7}$$

by the use of one-qudit unitary operations from the same sets.

Now we will present a more general protocol in which more than one sender can exploit the entanglement. For the pedagogical reason let us begin with a specific case when $d$ is a square of some integer $m$ and we have two senders ($N=2$). Then $n$ can be written in digital system of base $m$ as

$$n = (\alpha_1 \alpha_2)_m \qquad (8)$$

which means $n = \alpha_1 \cdot m + \alpha_2$ and $\alpha_1, \alpha_2 = 0,1,...,m-1$. If we choose $n_1 = \alpha_1 \cdot m$ and $n_2 = \alpha_2$ then a unique representation of $n$ is established. Simply speaking each sender can change one digit of $n$ (Eq. (8)). In this way each sender can transmit $1.5\log_2 d$ bits of information.

Now we will return to the general case. Let us express $d$ as a product of positive integers $p_k$ i.e.

$$d = p_1 p_2 ... p_N. \qquad (9)$$

It is clear that every integer $d$ can be written in this form (some of $p_k$ can be equal to 1). Let us define integers $q_k$ as

$$q_k = p_{k+1} p_{k+2} ... p_N \quad (k=1,2,...,N-1), \qquad (10)$$

$$q_N = 1.$$

With the use of the parameters $q_k$ we are able to present a systematic construction of the sets $S_k$. For each $k$ the set $S_k$ is defined as

$$S_k = \{n_k = \mu_k q_k \; : \; \mu_k = 0,1,...,p_k -1\}. \qquad (11)$$

This definition guarantees that each $0 \leq n \leq d-1$ can be written in a unique way as $n = \mu_1 q_1 + \mu_2 q_2 + ... + \mu_N q_N$. This equation can be understood as a representation of $n$ in the mixed base digital system [5]. We see that the number of elements of $S_k$ is $|S_k| = p_k$, so each sender is allowed to perform one of $p_k d$ unitary operations $U_{i_k}^{n_k}$. Thus, he can transmit $\log_2 p_k d$ bits of information. Because

$$\log_2 p_1 d + \log_2 p_2 d + \ldots + \log_2 p_N d = \log_2 p_1 p_2 \ldots p_N d^N = \log_2 d^{N+1} \qquad (12)$$

all senders together can transmit $\log_2 d^{N+1}$ bits of information.

It is worth noting that in a case when only $N'$ of $p_k \neq 1$ one can use only $N'+1$ particles in the entangled state and the product state of $N - N'$ particles. More explicitly, if $p_k \neq 1$ for $k = 1,2,\ldots,N'$ then the initial state can be taken in the form

$$|\Phi_{00\ldots 0}\rangle = \left( \sum_j \underbrace{|j\rangle \otimes \ldots \otimes |j\rangle}_{N'+1} \right) \otimes \underbrace{|0\rangle \otimes \ldots \otimes |0\rangle}_{N-N'} / \sqrt{d} . \qquad (13)$$

Let us present a simple example of dense coding among three parties in four dimensions ($N = 2, d = 4$). There are two possible choices for decomposition of $d$. The first one is $4 = 4 \times 1$ and the second one is $4 = 2 \times 2$. Let us consider the first case. We see that $p_1 = 4, p_2 = 1$ which leads to $S_1 = \{0,1,2,3\}, S_2 = \{0\}$. Thus, the first sender can transmit 4 and the second one only 2 bits of information. This is the case of the protocol of Liu, Long, Tong and Li [4]. The second decomposition gives $p_1 = 2, p_2 = 2$ and, consequently, $S_1 = \{0,2\}, S_2 = \{0,1\}$. So every sender can transmit 3 bits of information. This is an example of a fully symmetric scheme in which every sender can transmit the same amount of information. Generally fully symmetric dense coding scheme allowing every sender to transmit $x$ bits of information, requires qudits of the dimension $d = 2^{xN/(N+1)}$.

It should be mentioned that our protocol works even if one chooses $p_k$ satisfying the inequality $d > \delta = p_1 p_2 \ldots p_N$ instead of the equality $d = p_1 p_2 \ldots p_N$. In this case, instead of equation (12) we get

$$\log_2 p_1 d + \log_2 p_2 d + \ldots + \log_2 p_N d = \log_2 \delta\, d^N , \qquad (13)$$

which is less than $\log_2 d^{N+1}$ but is still more than $\log_2 d^N$. For example, in the case of dense coding among three parties in five dimensions ($N = 2, d = 5$) one can choose not only $p_1 = 5, p_2 = 1$ but also $p_1 = 2, p_2 = 2$. In the case of $p_1 = 5, p_2 = 1$ the first sender can transmit $\log_2 25 = 4.64$ and the second one $\log_2 5 = 2.32$ bits of information. In the case of $p_1 = 2, p_2 = 2$, every sender can transmit $\log_2 10 = 3.32$ bits of information.

In conclusion we have found a systematic way for constructing general schemes for high-dimensional multiparty superdense coding. We have also shown that in specific cases one can use $N'+1$ ($N' < N$) rather than $N+1$ particles in the maximally entangled state.

*Email address: agie@amu.edu.pl

**Email address: antwoj@amu.edu.pl